\documentclass[preprint,aps,showpacs,nofootinbib,preprintnumbers,amsmath,amssymb]{revtex4-1}
\usepackage{amssymb}
\usepackage{}
\usepackage{epsfig}
\usepackage{graphicx}% Include figure files
\usepackage{subfigure}
\usepackage{dcolumn}% Align table columns on decimal point
\usepackage{bm}% bold math
\usepackage[usenames ,dvipsnames]{xcolor}
\usepackage{slashed}

\begin{document}

\title{Imprint of Multi-component Dark Matter on AMS-02}

\author{Chao-Qiang~Geng$^{1,2,3,4}$\footnote{geng@phys.nthu.edu.tw},
Da~Huang$^{1,4}$\footnote{dahuang@phys.nthu.edu.tw}
and
Lu-Hsing Tsai$^{1,4}$\footnote{lhtsai@phys.nthu.edu.tw}}
  \affiliation{$^{1}$Department of Physics, National Tsing Hua University, Hsinchu, Taiwan\\
  $^{2}$Physics Division, National Center for Theoretical Sciences, Hsinchu, Taiwan\\
$^{3}$College of Mathematics \& Physics, Chongqing University of Posts \& Telecommunications, Chongqing, 400065, China\\
$^{4}$Kavli Institute for Theoretical Physics China (KITPC),
CAS,
%Chinese Academy of Science,
Beijing 100190 ,China}

\date{\today}
\begin{abstract}
The multi-component decaying dark matter (DM) scenario is investigated to explain the possible excesses in the positron fraction by PAMELA and recently confirmed by AMS-02, and in the total $e^+ +e^-$ flux observed by Fermi-LAT. By performing the $\chi^2$ fits, we find that two DM components are already enough to give a reasonable fit of both AMS-02 and Fermi-LAT data. The best-fitted results show that the heavier DM component with its mass 1.5 TeV dominantly decays through the $\mu$-channel, while the lighter one of 100 GeV mainly through the $\tau$-channel. As a byproduct, the fine structure around 100 GeV observed by AMS-02 and Fermi-LAT can be naturally explained by the dropping due to the lighter DM component. With the obtained model parameters by the fitting, we calculate the diffuse $\gamma$-ray emission spectrum in this two-component DM scenario, and find that it is consistent with the data measured by Fermi-LAT. We also construct a microscopic particle DM model to naturally realize the two-component DM scenario, and point out an interesting neutrino signal which is possible to be measured in the near future by IceCube.
\end{abstract}

%\pacs{95.35.+d, 98.70.Sa, 13.85.Tp, 14.80.-j}
%\keywords{}
\maketitle
\section{Introduction}
The constitution of the cosmic ray (CR) can always tell us a lot about our Galaxy and our universe. Recently, the AMS-02 collaboration has published the first measurement of the positron fraction $e^+/(e^- + e^+)$ in CR with a high precision, which shows a continuous rise from $5$ up to $350$ GeV~\cite{AMS02} and confirms the general behavior previously measured by CAPRICE~\cite{Boezio:2000zz}, HEAT~\cite{DuVernois:2001bb, Beatty:2004cy}, AMS-01~\cite{Aguilar:2007yf}, PAMELA~\cite{PAMELA, PAMELA2} and Fermi-LAT~\cite{FermiLAT:2011ab}. The observed uprise is in stark contrast with the conventional expectations based on the secondary-origin positrons, whose fraction is just monotonically decreasing with energy. Furthermore, the total flux spectrum of electrons and positrons measured by ATIC~\cite{Chang:2008aa}, PPB-BETS~\cite{PPB}, HESS~\cite{HESS1,HESS2}, Fermi-LAT~\cite{Ackermann:2010ij,FermiLAT} and more recently by AMS-02 is harder than those expected from the conventional astrophysical background, indicating some excesses in the energy range higher than 10 GeV.
All these results imply that there exist some extra exotic $e^{\pm}$ sources in our Galaxy which are unknown to us.

In the literature, there have been many possible mechanisms, such as the astrophysical source like pulsars~\cite{pulsar}, annihilating dark matter (DM)~\cite{DMindependent, annihilation, annihilationAMS, AnnihilationDecay} and decaying DM~\cite{DMindependent, decay, decayAMS, 3bodydecay, 3bodydecayAMS,Ishiwata:2009vx}. However, it is pointed out in Refs.~\cite{Chen:2009gd, Feng:2013zca, 2body, 2bodyAMSa, 2bodyAMSb} that there is a tension between the AMS-02 positron fraction and the Fermi-LAT total flux since the slope of the former decreases one order of magnitude from 10 to 250 GeV~\cite{AMS02}, while the latter is much flatter. In particular, for the simplest scenario with a single type of DM whose decay is mainly through the leptonic two-body channels, it is difficult in performing a good fit of the AMS-02 and Fermi-LAT data simultaneously~\cite{Feng:2013zca,2bodyAMSb}. In order to reduce this tension, we need to resort to some more complicated models, such as three/four-body decaying/annihilating~\cite{3bodydecayAMS}, asymmetric decaying~\cite{asymmetricDM}, dynamical DMs~\cite{dynamicalDM} as well as other astrophysical solutions~\cite{pulsar,Feng:2013zca}.

More interestingly, the positron fraction from AMS-02 and the total $e^+ + e^-$ flux from Fermi-LAT and many other experiments show a structure with ``flash damp" or ``jerk" around 100 GeV. Since this fine structure is observed in more than two independent experiments, we think it is reasonable to take it seriously, though it can also be caused by the statistical fluctuations.

In this paper, we propose a multi-component decaying DM scenario with two-body leptonic decay channels. Such a scenario can reconcile the tension between the AMS-02 and Fermi-LAT data since the change of the slope in the spectrum is achieved by the different channels of two DM components. Moreover, the fine structure in the two data sets has the natural explanation that the lighter DM drops around 100 GeV. Although the similar scenario has already been considered in Ref.~\cite{multiDM}, our present discussion is more general.

As mentioned in Ref.~\cite{Cirelli:2012ut}, the most stringent constraint on the decaying DM models comes from the cosmic diffuse $\gamma$-rays, which was precisely measured by EGRET~\cite{EGRET} and more recently by Fermi-LAT~\cite{FermiLAT_Gamma}. In our present work, the dark matter contribution to the diffuse $\gamma$-rays could be produced from the leptonic final state radiation associated to the DM leptonic decay and the scattering of the resultant electrons/positrons to the interstellar medium (ISM) via bremsstrahlung as well as the low-energy photons inside and outside of our Galaxy via the inverse Compton (IC) process. As a result, with the parameters obtained by fitting the AMS-02 positron fraction and the Fermi-LAT total $e^+ + e^-$ flux, the total diffuse $\gamma$-ray spectrum is completely fixed. We will demonstrate that the predicted diffuse $\gamma$-ray spectrum does not exceed the Fermi-LAT bound, and somehow agrees with the measured spectrum well.

The paper is organized as follows. In Sec.~\ref{pst_fit}, we first perform the $\chi^2$-fitting of the AMS-02 positron fraction and the Fermi-LAT total electron/positron flux with a single component DM. We then propose the multi-component DM scenario to fit the data by carefully examining the simplest case with only two components. In Sec.~\ref{gamma}, we predict the total diffuse $\gamma$-ray spectrum and compare it with the Fermi-LAT data. In Sec.~\ref{model}, we build a simple microscopic model to realize the two-component DM scenario. Our conclusions are presented in Sec.~\ref{conclusion}.

\section{Fitting Decaying Dark Matter Models with AMS-02 and Fermi-LAT data }\label{pst_fit}
\subsection{Sources and Propagation of Cosmic-Ray in the Galaxy}
The propagation of various charged CR particles in our Galaxy is well described by the general diffusion-reacceleration equation, given by~\cite{DiffuseEquation}
\begin{eqnarray}\label{diffusionEq}
\frac{\partial \psi}{\partial t} &= & Q({\bf x},p) + \nabla \cdot(D_{xx}\nabla\psi-{\bf V}_c\psi)+\frac{\partial}{\partial p}p^2 D_{pp} \frac{\partial}{\partial p}\frac{1}{p^2}\psi-\frac{\partial}{\partial p}\big[ \dot{p}\psi - \frac{p}{3}(\nabla \cdot {\bf V}_c )\psi \big]\nonumber\\
&& -\frac{1}{\tau_f}\psi-\frac{1}{\tau_r}\psi,
\end{eqnarray}
where $\psi({\bf x}, p, t)$ is the number density of CR particles per unit of momentum,  $Q({\bf x}, p)$ is the source term, and $D_{xx}$ is the spatial diffusion coefficient which is parameterized as a power law $D_{xx} = \beta D_0(\rho/\rho_r)^{\delta}$ with $\rho = p/(Ze)$ the rigidity of the cosmic ray, $\rho_r$ the reference rigidity, $\beta = v/c$ the velocity and $\delta$ the power spectral index. The normalization constant $D_0$ and the power index $\delta$ are determined by fitting the experimental values of the secondary-to-primary ratios, such as $\mbox{B/C}$, and the unstable-to-stable ratios of secondary particles, such as ${}^{10}\mbox{Be}/{}^{9}\mbox{Be}$ and ${}^{26}\mbox{Al}/{}^{27}\mbox{Al}$. The overall convection driven by the stellar wind is characterized by the convection velocity ${\bf V}_c$, and the reacceleration process is described by the diffusion coefficient in the momentum space $D_{pp}=4 V_a^2 p^2/(3D_{xx}\delta(4-\delta^2)(4-\delta))$. In Eq.(\ref{diffusionEq}), $\dot{p}=d p/d t$ denotes the momentum loss rate, while $\tau_f$ and $\tau_r$ the time scales for the nuclei fragmentation and radioactive decay, respectively. In the usual CR propagation model, the CR diffusion is confined in a Galactic halo which is parametrized as a cylinder with half-height $z_h$ and radius $r_h$, while the densities of the CR components vanish at the boundary of the halo. In our computation, we take $z_h=4$~kpc and $r_h=20$~kpc.

The source term $Q({\bf x},p)$ for the primary particles is the product of the particle injection spectra $q^{n,e}(\rho)$ and the CR source spatial distribution $f(R,z)$, which are broken power-law functions with respect to the rigidity $\rho$ and the supernova-remnant~(SNR) type:
\begin{eqnarray}
q^{n,e}(\rho) &\propto& \Big(\frac{\rho}{\rho_{br}^{n,e}}\Big)^{-\gamma^{n,e}_1(\gamma^{n,e}_2)},\\
f(R,z) &\propto& \Big( \frac{R}{R_\odot} \Big)^a \exp\Big[ -\frac{b(R-R_\odot)}{R_\odot} \Big]\exp\Big(-\frac{|z|}{z_s}\Big),
\end{eqnarray}
respectively, where $\gamma^{n,e}_{1(2)}$ are the spectral index below (above) the nucleus and electron broken rigidities $\rho_{br}^{n,e}$ , $R_\odot = 8.5$~kpc is the distance between the Galactic center and our solar system and $z_s = 0.2$~kpc is the characteristic height of the Galactic disk. Here, we have adopted $a=1.25$ and $b=3.56$ by following Ref.~\cite{Trotta:2010mx}.

The collisions of the primary CR particles in the interstellar medium (ISM) would produce the secondary particles. For our present interest, the secondary positrons and electrons are the final   products of the decay chain of the pions and kaons originated from such collisions, which can be calculated along with solving the CR diffusion equations.

The primary electrons and secondary electrons/positrons constitute the background of the $e^+ +e^-$ flux. However, in order to explain the AMS-02 and Fermi-LAT results, we need to introduce additional primary source terms $Q^{\rm DM}_{\pm}$ into the positron/electron diffusion equations. In this study, we shall always adopt the isothermal profile as our DM distribution in the Galaxy~\cite{isothermal}, which is given by:
\begin{eqnarray}
\rho(r)=\rho_0{r_c^2+r_\odot^2 \over r_c^2+r^2}\;,
\end{eqnarray}
where $\rho_0=0.43~{\rm GeV}\cdot{\rm cm}^{-3}$, $r_c=2.8~{\rm kpc}$ is the core radius, and $r_\odot=8.5~{\rm kpc}$ is the distance between the galactic center and our solar system as $R_\odot$. The variable $r$ is the distance from the Galactic center to the position of the DM source. The $e^+/e^-$ injection spectra induced by the DM decays are, however, much model-dependent, so that they are introduced in the corresponding subsections below.

After the propagation of the CR by taking into account the energy losses for electrons/positrons by ionization, Coulomb interaction, inverse Compton (IC) scattering, bremsstrahlung and synchrotron radiation under the galactic magnetic fields, we can obtain the electron/positron fluxes observed around the earth through the relation $\Phi_{e}=(c/ 4\pi)\psi(E)$. In the present work, we use the numerical package {\sc GALPROP}~\cite{GALPROP} to consistently solve the coupled diffusion-reacceleration equations for various CR components by including the $e^+/e^-$ contribution from the decaying DM sources. For our numerical calculations, we apply the parameter set as shown in Table~\ref{parameters}.
\begin{table}
\caption{The parameters for the diffuse propagation, primary electron, and primary proton.}
\begin{tabular}[t]{|cccc|ccc|ccc|}
\hline
\multicolumn{4}{|c|}{diffuse coefficient}&\multicolumn{3}{c|}{primary electron}
&\multicolumn{3}{c|}{primary proton} \\
\hline
$D_0(\mathrm{cm}^2\mathrm{s}^{-1})$ & $\rho_r(\rm{MV})$  &$\delta$ &$v_A({\rm km\, s}^{-1})$&$\rho_{\rm br}^e(\rm{MV})$&$\gamma_1^e$&$\gamma_2^e$
&$\rho_{\rm br}^p(\rm{MV})$&$\gamma_1^n$&$\gamma_2^n$\\
\hline
$5.3\times10^{28}$& $4.0\times10^{3}$  & 0.33&$33.5$
&$4.0\times10^{3}$ &1.54&  2.6 &$11.5\times10^{3}$ &1.88 &2.39 \\
\hline
\end{tabular}\label{parameters}
\end{table}
As a result, the total fluxes $\Phi^{\rm (tot)}_{e,p}$ for electrons and positrons can be expressed by
\begin{eqnarray}
\Phi^{\rm (tot)}_{e}&=&\kappa \Phi^{\rm (primary)}_{e}+\Phi^{\rm (secondary)}_{e}
+\Phi^{\rm DM}_{e}\;,\nonumber\\
\Phi^{\rm (tot)}_{p}&=&\Phi^{\rm (secondary)}_{p}
+\Phi^{\rm DM}_{p}\;,
\end{eqnarray}
where the factor $\kappa$ is inserted to account for the uncertainty in the normalization for the primary electron flux, which should be fixed with other parameters of the model in the fitting procedure.

Finally, due to the solar winds and the heliospheric magnetic field at the top of the atmosphere (TOA) of the earth, the fluxes of the CR particles would be affected. Here, we use the simple force-field approximation~\cite{Gleeson:1968zza} to account for this solar modulation effect, that is, the measured electron/positron fluxes at the TOA are related to the interstellar ones via:
\begin{eqnarray}
\Phi^{\rm TOA}_{e/p} (T_{\rm TOA}) = \Big(\frac{2 m_e T_{\rm TOA}+T^2_{\rm TOA}}{2m_e T+T^2}\Big) \Phi^{\rm tot}_{e/p},
\end{eqnarray}
where $T_{\rm TOA}=T-\phi_F$ is the kinetic energy of the electrons/positrons at the top of the atmosphere and numerically we take the potential $\phi_F = 0.55$~GV.

\subsection{General Discussion of Decaying Dark Matter Scenario}\label{general}
\begin{figure}
\centering
\includegraphics[width=8cm]{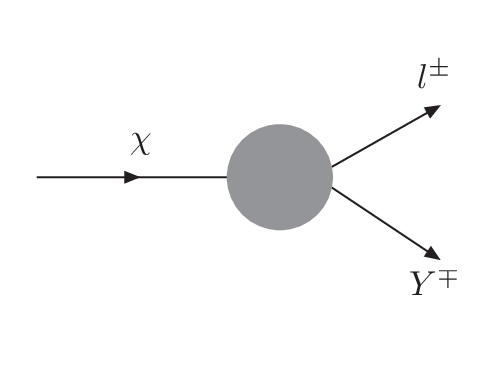}
\caption{Illustration for the process of the DM particle $\chi$ decaying into charged leptons $\l^\pm$ and heavy charged particles $Y^{\mp}$.}\label{Fig_DMDecay}
\end{figure}

In the decaying DM scenario, although the lifetime of each DM component is typically of ${\cal O}(10^{26}{\rm s})$ which is much longer than the age of the universe $\tau_U \approx 4\times 10^{17}$~s, it is remarkable that such a low decay rate is already enough to provide a sufficient amount of positrons and/or electrons to explain the AMS-02 and Fermi-LAT excesses. The $e^\pm$ source terms $Q^{\rm DM}_{e,p}$ induced by the DM decays can be generally expressed as
\begin{eqnarray}
Q({\mathbf x},p)^{\rm DM}_{e,p}= \sum_i{\rho_i(\mathbf x)\over \tau_i M_i} \Big( \frac{dN_{e,p}}{dE} \Big)_i \;,
\end{eqnarray}
where $M_i$, $\tau_i$ and $\rho_i(\mathbf x)$ denote the mass, lifetime and energy density distribution for the $i$-th DM component in our Galaxy, respectively, and $(dN_{e,p}/ dE)_i$ is the differential electron/positron multiplicity per annihilation, which depends on the main decaying processes of the DM. In the following, we focus on the scenario in which all DMs dominantly decay through two-body leptonic processes $\chi_i\rightarrow  l^\pm Y^\mp$, where $\chi_i$ represents the $i$-th DM particle, $l=e$, $\mu$ and $\tau$, and $Y$ is another heavy charged particle with mass $M_Y$, which is illustrated in Fig.~\ref{Fig_DMDecay}.
In this simple two-body decay scenario, $(dN_{e,p}/ dE)_i$ is fully determined by the kinematical analysis, in which the produced leptons have a definite energy $E_c$ which can be written as a function of $M_i$ and $M_Y$. Thus, when $l=e$, the $e^+/e^-$ energy spectrum is just a delta function:
\begin{equation}
\Big({dN^e\over dE}\Big)_i= {1\over E_{ci}}\delta(1-x)\;,\;
\end{equation}
where $x=E/E_{ci}$. For the $l=\mu$ and $\tau$ cases, the subsequent decay would give one or more positrons/electrons. The normalized $e^{\pm}$ energy spectrum $d{N^{\mu }}/dE$ has the following analytical expression:
\begin{eqnarray}
\Big({dN^\mu\over dE}\Big)_i&=&{1\over E_{ci}}[3(1-x^2)-{4\over3}(1-x)]\theta(1-x)\;,\;
\end{eqnarray}
with $x=E/E_{ci}$, while $dN^\tau/dE$ can be obtained by the simulation of the $\tau$ decay with {\sc PYTHIA}~\cite{PYTHIA}. For the general situation with all decay channels of $l=e,\mu$ and $\tau$ simultaneously, the total electron/positron energy distribution from the decaying DMs can be normalized as
\begin{eqnarray}\label{Norm_Spectrum}
\Big({dN_{e,p}\over dE}\Big)_i={1\over2}\Big[\epsilon^e_i\Big({dN^e\over dE}\Big)_i+\epsilon^\mu_i\Big({dN^\mu\over dE}\Big)_i+\epsilon^\tau_i\Big({dN^\tau\over dE}\Big)_i\Big]\;,
\end{eqnarray}
where $\epsilon^{e,\mu,\tau}_i$ are the branching ratios for three leptonic channels of the $i$-th DM with the relation $\epsilon^{e}_i+\epsilon^{\mu}_i+\epsilon^{\tau}_i=1$, and the factor $1/2$ takes into account that $e^-$ and $e^+$ come from two different channels. This normalization relation means that the leptonic decay channels dominate over other ones, realizing the leptophilic scenario which is favored by the current measurement of the  antiproton flux spectrum in CR by PAMELA~\cite{Adriani:2010rc}. These branching ratio parameters will be determined by fitting the $e^\pm$ spectra in the following subsections.

Note that in our present setup, we assume that the DM decays will give the same amount of electrons and positrons, rather than the asymmetric DM scenario~\cite{asymmetricDM}. Moreover, at the first sight our present leptonic decay channels are different from the ones in ${\rm DM}\rightarrow l^+ l^-$ usually considered in the literature, but the final $e^\pm$ spectra are essentially identical just by the replacement of the energy cutoff $E_{ci}$ with the half DM mass $M_i/2$. Clearly, with the simple rescaling of the obtained DM lifetimes $\tau_i$ and masses $M_i$, our fitting may also be generalized to ${\rm DM}\rightarrow l^+ l^-$.

\subsection{Fitting Results with the Single-Component Decaying Dark Matter}
In this subsection, we shall concentrate on the simplest case with only one DM component. Within the above general framework with the DM mass $M=3030$~GeV, we have 5 parameters: the primary electron spectrum normalization factor $\kappa$, energy cutoff $E_c$, DM lifetime $\tau$ and two independent branching ratios $\epsilon^e$ and $\epsilon^\tau$,  which will be determined by fitting the data points of the AMS-02 positron fraction and the Fermi-LAT total $e^\pm$ flux. In addition, $\epsilon^e$ and $\epsilon^\tau$ should be subject to the constraint $\epsilon^e + \epsilon^\tau \leq 1$ by considering the possible contribution from the $\mu$-channel. Note that the Fermi-LAT data show that the $e^\pm$ excess extends as high as to 1 TeV, which indicates that the DM cutoff $E_c$ should be, at least, equal to or larger than 1 TeV. Since our purpose is to discuss the generic feature of the proposed decaying DM scenario, we fix $E_c$ to be 1 TeV, 1.3 TeV and 1.5 TeV respectively, while fitting other four parameters. In this work, we take the 42 data points of the positron fraction from AMS-02~\cite{AMS02} with energy above 10 GeV and the 26 data points of the total flux of electrons and positrons from Fermi-LAT~\cite{FermiLAT}. The selection constraint with energy above 10 GeV is set in order to reduce the effects of the solar modulation. In total, we consider 68 data points in our global fits. For the fitting procedure, we use the simple $\chi^2$-minimization method, in which the $\chi^2$-function is constructed as
\begin{eqnarray}
\chi^2=\sum_{i=1}^{68}\Big({y^{\rm th}_i-y^{\rm exp}_i\over \sigma_i}\Big)^2\;,
\end{eqnarray}
where $y^{\rm th}_i$ are the theoretical predictions for the positron fraction or the total $e^+ + e^-$ flux and $y_i^{\rm exp}$ are the corresponding experimental data points with errors $\sigma_i$. The index $i$ runs over all the data points. The point in the parameter space which gives the minimal $\chi^2$ value will be the best-fit point for our DM model.
\begin{table}
\caption{Points of the parameter space for different cutoff values of $E_c$, which lead to minimal $\chi^2$ where the DM mass is chosen as {$M=3030$ GeV}. }
\begin{tabular}[t]{lccccccc}
\hline
$E_c$(GeV) & $\kappa$  &$\epsilon^{e}$ &$\epsilon^{\mu}$&$\epsilon^{\tau}$  & $\tau(10^{26}{\rm s})$ & $\chi_{\rm min}^2$& $\chi_{\rm min}^2/d.o.f.$\\
\hline
1000& 0.73  & 0.09 & 0&  0.91 & {0.66}  &463&7.35\\
1300& 0.72  & 0.04 & 0&  0.96 & {0.71}  &516&8.19\\
1500& 0.71  & 0.02 & 0&  0.98 & {0.74}  &541&8.46\\
\hline
\end{tabular}
\label{tab_1DM}
\end{table}

\begin{figure}
\centering
\includegraphics[width=16cm]{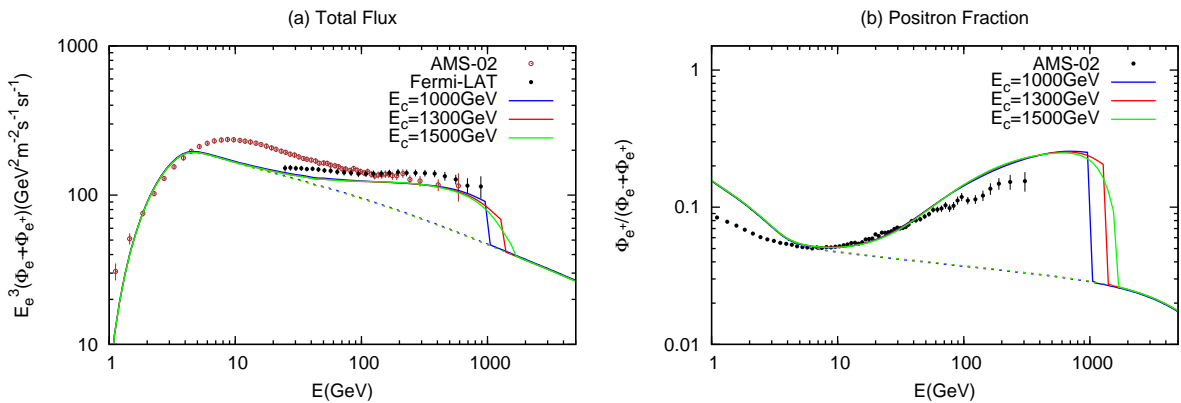}
\caption{(a) Total flux and (b) positron fraction from the DM contributions with the best-fitting parameters given in Table~\ref{tab_1DM}.
%, {\color{red} where the black and red dots in (a) represent the  total flux data from Fermi-LAT and  AMS-02, respectively}
}\label{Fig_1DM}
\end{figure}

The best-fit results are shown in Table~\ref{tab_1DM} and Fig.~\ref{Fig_1DM} for the three cases with the electron energy cutoff at $E_c=1$ TeV, 1.3 TeV and 1.5 TeV, respectively. From Table~\ref{tab_1DM}, we find that $\chi_{\rm min}^2/{\rm d.o.f}$ is always larger than 7 and it tends to increase with a larger $E_c$, suggesting that the single DM models with the five parameters $\{\kappa, E_c, \tau, \epsilon^e,\epsilon^\tau \}$ cannot provide a reasonable fit to the AMS-02+Fermi-LAT data. This result agrees with the previous studies in the single-component decaying DM models~\cite{2bodyAMSa}. It is also interesting to note that for all the three cases, the best fits indicate that the DM $\mu$-channel does not contribute the electron/positron flux since $\epsilon^{e,\tau}$ always saturate the constraint $\epsilon^e+\epsilon^\tau =1$.

From the technical perspective, the failure of the fitting can be attributed to the fact that the positron fraction from all three leptonic DM decaying channels are harder than the measured spectrum by AMS-02. Since the AMS-02 data in the low energy range, around $E \simeq {\rm 10}$~GeV, have very small errors and thus dominate the value of $\chi^2$, the parameters are, in fact, already fixed by saturating those data points. The resultant $e^\pm$ spectra deviate from the Fermi-LAT and AMS-02 data at high energies as depicted in Fig.~\ref{Fig_1DM}.

Therefore, the insufficiency to fit the AMS-02 positron fraction and the Fermi-LAT total $e^+ + e^-$ flux implies that the single-decaying DM scenario should be extended to a more complicated situation. There are several ways to do this. One interesting idea is to split the whole DM density into multiple components, which will be discussed in a great detail in the next subsection.

\subsection{Fitting with the Two-Component Decaying Dark Matter}
The multiple-component DM scenario is very interesting from a phenomenological perspective~\cite{multiDM,DoubleDiskDM,dynamicalDM}. In this subsection, we consider the implication of the multiple-component DM to the indirect DM search. In particular, we shall show that with the two DM components, denoted by ${\rm DM}_{L(H)}$, representing the lighter (heavier) DM, it is enough to accommodate the AMS-02 positron fraction and the Fermi-LAT total positron/electron flux simultaneously. In addition, if the decay of the ${\rm DM}_L$ to $e^\pm$ terminates at around 100 GeV, the cutoff energy $E_{cL}$ would manifest itself as the fine structure of ``jerk" or ``flash damp"~\cite{multiDM} in the positron fraction and total $e^+ + e^-$ flux spectra, as implied by AMS-02, Fermi-LAT, and many others.

For simplicity, we shall assume that each of the two components carries half of the total energy density of DM in the Galaxy and in the whole universe. We also take that the three two-body charged leptonic decay channels are the major decaying processes for both DM components as specified in Sec.~\ref{general}. Consequently, the extra electron/positron source term due to DM decays should be modified to:
\begin{eqnarray}
Q({\mathbf x},p)^{\rm DM}_{e,p}={\rho({\mathbf x})\over 2}\Big[{1\over\tau_L M_L}\Big({\frac{dN_{e,p}}{dE}}\Big)_L
+{1\over\tau_H M_H}\Big({dN_{e,p}\over dE}\Big)_H\Big]\;,
\end{eqnarray}
where the subscripts $L$ and $H$ represent the quantities corresponding to $\mathrm{DM}_L$ and $\mathrm{DM}_H$ with choosing $M_{L,H}=416$ and 3030 GeV, respectively. In order to make the total flux excess cover the whole Fermi-LAT energy range, the energy cutoff for ${\rm DM}_H$ is set to $E_{cH}=1500~\mathrm{GeV}$. The fine structure around 100 GeV shown in the AMS-02 data determines $E_{cL}=100~{\rm GeV}$. The normalized total electron/positron differential multiplicity $\Big({dN_{e,p}\over dE}\Big)_{H(L)}$ for each DM is defined in Eq.~(\ref{Norm_Spectrum}). Hence, in the present two-component DM model, we are left with 7 parameters needed to be fixed by the $\chi^2$ fitting: $\tau_i$, $\epsilon^e_i$ and $\epsilon^\tau_i$ for each DM, together with the primary electron normalization uncertainty $\kappa$. The data selection and fitting procedure are the same as in the single-component DM scenario.

After some tentative fits, we find that the minimum of $\chi^2$ can be obtained when $\epsilon^{e}_H=\epsilon^{\tau}_H=0$, and $\epsilon^e_L+\epsilon^{\tau}_L=1$ which implies $\epsilon^\mu_L=0$ . Thus, in order to enhance the accuracy and stability of the fit, we turn off these three channels by requiring $\epsilon^e_H=\epsilon^\tau_H=0$ and $\epsilon^\tau_L=1-\epsilon^e_L$, and fit the rest four parameters again. The value of $\chi_{\rm min}^2$ and its corresponding parameters are given in Table~\ref{tab_2DM}.
\begin{table}
\caption{The point in the parameter space which gives the minimal value of $\chi^2$ with the DM masses and cutoff energies taken as {$M_{L,H}=(416,3030)~{\rm GeV}$} and $E_{cL,H}= (100,1500)~{\rm GeV}$, respectively.}
\begin{tabular}[t]{l|cccc|cccc|cc}
\hline
 $\kappa$ & $\epsilon^{e}_H$& $\epsilon^{\mu}_H$  & $\epsilon^{\tau}_H$
 & $\tau_H(10^{26}{\rm s})$& $\epsilon^{e}_L$& $\epsilon^{\mu}_L$ & $\epsilon^{\tau}_L$
&$\tau_L(10^{26}{\rm s})$& $\chi_{\rm min}^2$& $\chi_{\rm min}^2/d.o.f.$\\
\hline
 0.844 & 0  &1 &  0 & {0.76} & 0.018  &0 &  0.982 & {0.82}  & 62.3& 1.06\\
\hline
\end{tabular}
\label{tab_2DM}
\end{table}
The minimum of $\chi^2/{\rm d.o.f}$ is only 1.06, representing the goodness of the fitting.   The best-fit results tell us that ${\rm DM}_H$ decays only through the muon-channel while the lighter one mainly through the electron and tau channels with the latter being the dominant one. The determination of the flavor dependence of the DM decay channels displays the power of the indirect DM search method.

\begin{figure}
\centering
\includegraphics[width=16cm]{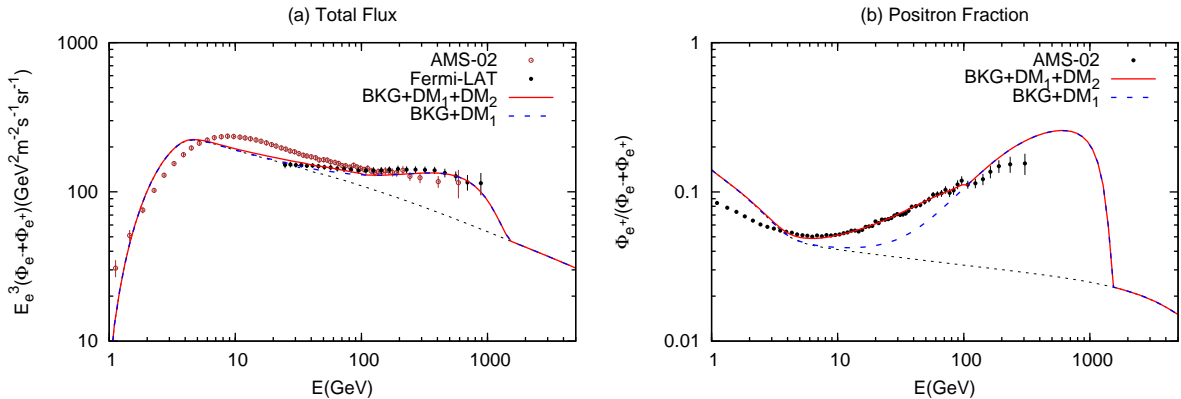}
\caption{(a) Total flux and (b) positron fraction from the DM contributions with the parameters given in Table~\ref{tab_2DM}. 
%{\color{red} In Fig.~(a), the black dots represent the Fermi-LAT total flux data, while the red ones the recent AMS-02 data.} 
}\label{Fig_2DM}
\end{figure}

The predicted positron fraction and total $e^+ + e^-$ flux based on the best-fit parameters are depicted in Fig.~\ref{Fig_2DM}.
The fine structure around 100 GeV is evident in the positron fraction, and less significant in the total $e^+ + e^-$ flux, 
both of which agree well with the experimental data of AMS-02 and Fermi-LAT. If this fine structure persists and 
becomes clearer when more data are accumulated by AMS-02, Fermi-LAT and many others in the near future, 
it would be an important support to the multi-component DM scenario.

Finally, we would like to give some discussions on our fitting results when the data sets are changed. 
More recently, the AMS-02 collaboration has showed their measurement on the total $e^+ + e^-$ flux. 
If we use the AMS-02 data on the total flux instead of the Fermi-LAT ones, we have checked that the present setup 
would give essentially the same level of the fit. The main difference is that the ratio of the primary electron source 
is increased and the lifetime of the lighter DM is reduced, reflecting the fact that we need more $e^+/e^-$ at the low energy 
shown in the AMS-02 data. Another issue is related to our data-taking criterion that we only adopt the data point 
with the energy above 10 GeV. Since the error bars at the low energy are small, it is expected that the data just above 
this energy cutoff would give the most important statistical power so that the variation of this cutoff slightly 
would alter the final fitting result. However, 
%as has been checked with 
from the $\chi^2-$fitting, we find that the rise of the cutoff to 20 GeV does not lead to a large effect, 
which may be related to the fact that the precision of the AMS-02 positron fraction data in the whole energy range 
is much higher than that of the Fermi-LAT total $e^+ + e^-$ flux and thus still dominates the fitting.
% no matter what the low energy bound is chosen to be.}

\section{Diffuse Gamma Ray from Dark Matter Decay}\label{gamma}
As discussed in Refs.~\cite{Cirelli:2012ut, Papucci:2009gd, Beacom:2004pe, Gamma_DM, Gamma_extraDMIC, Gamma_extra, Ibarra:2007wg,Ishiwata:2009vx}, the positrons/electrons from the DM decays or annihilations are always accompanied with the emissions of high energy photons, which would contribute to the diffuse $\gamma$-ray background. In order for the decay of a typical DM candidate to account for the observed positrons and electrons in AMS-02 and Fermi-LAT, the associated flux of high-energy $\gamma$-rays would have the potential to exceed the diffuse $\gamma$-ray data by Fermi-LAT~\cite{FermiLAT_Gamma} and EGRET~\cite{EGRET}. As pointed in Refs.~\cite{Gamma_DM, Gamma_extraDMIC, Gamma_extra}, especially Ref.~\cite{Cirelli:2012ut, Papucci:2009gd}, a large range of the parameter space of the decaying DM models with the usual decay channels trying to explain the PAMELA and Fermi-LAT positron/electron excesses has already been excluded. Thus, it is necessary to consider whether our multi-component decaying DM scenario, in particular the two-component DM case discussed in the previous section, is still viable under the diffuse $\gamma$-ray constraints.

\begin{figure}
\centering
\includegraphics[width=12cm]{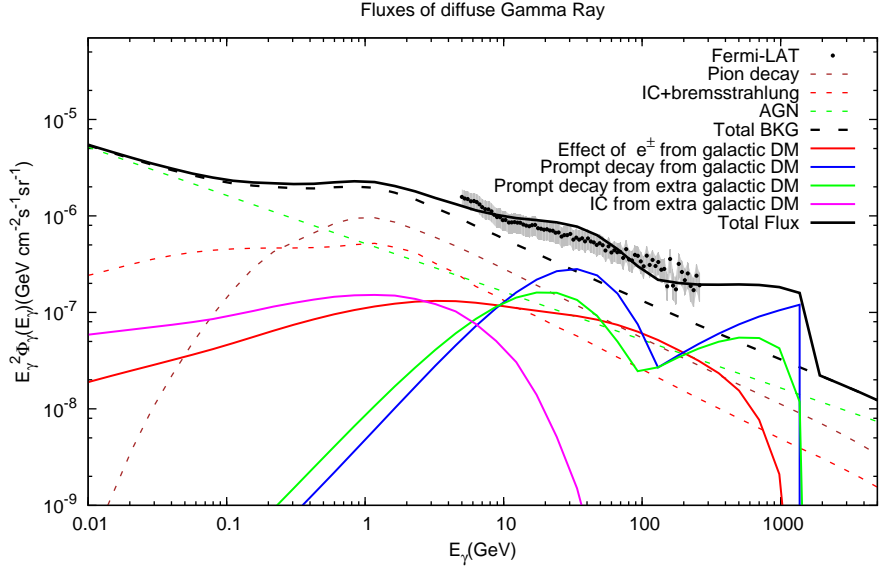}
\caption{Photon fluxes as a function of $E_\gamma$, where the black solid line is the total contribution from ordinary sources and DM, 
while the other solid lines are related to different parts of DM contributions, and the dashed lines correspond to ordinary sources inside 
or outside our galaxy. 
%{\color{red} 
Note that the gray band represents the total error, including all the systematic and statistical ones.
%}
}\label{Fig_GammaRay}
\end{figure}

In this section, we compute the total diffuse $\gamma$-ray flux by taking account of all possible $\gamma$-ray sources, including the usual astrophysical diffuse background $\gamma$-ray radiation inside and outside our Galaxy as well as the DM contributions. We will compare our result with the Fermi-LAT inclusive continuum photon spectrum~\cite{FermiLAT_Gamma}, which was measured within the energy range $4.8~{\rm GeV}<E_\gamma<264~{\rm GeV}$ for the sky in the high latitude with $|b|>10^\circ$ plus the Galactic center (GC) with $|b|<10^\circ$, $l<10^\circ$ and $l>350^\circ$. The conventional astrophysical background can be further divided into two parts: inside and outside the Galaxy. For the background $\gamma$-radiation, we have included three sources: pion decay, inverse Compton (IC) scattering, and bremsstrahlung, all of which are originated from the collisions of CR particles with the galactic interstellar medium (ISM) and low energy photons during the CR diffuse process. We use the {\rm GALPROP} code to numerically calculate the spectra of these three parts of high energy photons with the same numerical values of the CR diffusion parameters when we fit the two-component DM model with the AMS-02 and Fermi-LAT excesses as well as the primary electron normalization $\kappa=0.844$, obtained by the $\chi^2$-fitting in the last section. The extragalactic $\gamma$-ray background (EGB) is usually considered to be the superposition of contributions from unresolved extragalactic sources, such as the active galactic nuclei (AGN). In the present work, we adopt the following parameterization:
\begin{eqnarray}{
E^2\Phi_\gamma(E)=5.18\times 10^{-7}E^{-0.499}( {\rm GeV} {\rm cm}^{-2} {\rm sr}^{-1} {\rm s}^{-1})\;,}\label{Eq_ExtraSource}
\end{eqnarray}
which is obtained by fitting the low energy spectrum of the EGRET $\gamma$-ray data~\cite{EGRET,Ishiwata:2009vx}. The total sum of these two backgrounds is shown as the black dashed line in Fig.\ref{Fig_GammaRay}.

The DM decays can provide a lot of new sources for the $\gamma$-ray flux. Inside the Galaxy, the extra high-energy electrons/positrons as the decay products of the two DM components can induce the $\gamma$-rays by the collision with the ISM via bremsstrahlung and the scattering with the starlight, IR photons and the Cosmic Microwave Background~(CMB) via IC, both of which can also be computed by using the GALPROP package. Furthermore, we should also consider the $\gamma$-rays coming from the associated DM prompt decays. Since our two DM components decays involve $e,\mu$ and $\tau$ channels, $\gamma$-rays can be emitted via the internal bremsstrahlung~\cite{Beacom:2004pe} or final state radiation (FSR)~\cite{Gamma_DM, FermiLAT_Gamma} from the external lepton legs. For the $\mu$-channel, we also include the effects of the radiative muon decays~\cite{Gamma_DM}, {\emph e.g.}, $\mu^{+} \to e^{+} \bar{\nu}_\mu \nu_e\gamma$ and $\mu^{-} \to e^{-} \bar{\nu}_\mu \nu_e\gamma$. For the $\tau$-channel, the produced $\tau$ decays produce many $\pi^0$, which can further decay into two photons with the total spectrum parameterized as~\cite{Gamma_DM, Fornengo:2004kj}:
\begin{eqnarray}
\frac{d N_\gamma}{d y} = y^{-1.31} (6.94y-4.93y^2-0.51y^3)e^{-4.53y},
\end{eqnarray}
where $y=E_\gamma/M_{H,L}$ for two DM components. The three lepton FSRs, the radiative muon decays and the pion decays from $\tau$ will be called the prompt decay below. The relative size of these contributions can be completely determined by the fitted $\epsilon^{e,\mu,\tau}_{H,L}$ listed in Table~\ref{tab_2DM}.

Outside the Galaxy, the DM-induced $\gamma$-rays are mainly generated by the prompt decays and the ICs of the electrons and positrons from the DM decays with the CMB photons. Different from the prompt decays inside the Galaxy, we need to consider the $\gamma$-ray redshift effects caused by the cosmic expansion, which is encoded in the following formula~\cite{Gamma_extra}:
\begin{eqnarray}
\Big[E_\gamma^2 \frac{d \Phi_\gamma}{dE_\gamma}\Big]_{\rm eg} &=& E_\gamma^2\cdot \frac{c\, \Omega_{\rm DM}\rho_c}{4\pi H_0 \Omega_M^{1/2}}\int^\infty_1 dy \frac{y^{-3/2}}{\sqrt{1+\Omega_\Lambda/\Omega_M y^{-3}}}\cdot \nonumber\\
&& \frac{1}{2}\Big[\frac{1}{\tau_H M_H}\Big(\frac{d N_\gamma}{d(y E_\gamma)}\Big)_H+\frac{1}{\tau_L M_L} \Big(\frac{d N_\gamma}{d(y E_\gamma)}\Big)_L\Big]\, ,
\end{eqnarray}
where $y=1+z$ with $z$ being the redshift, $c$ is the speed of light, $H_0$ represents the present value of the Hubble parameter, $\rho_c$ is the critical density, and $(\Omega_{\rm DM},\Omega_M,\Omega_\Lambda)= (0.11889 h^{-2}, 0.14105 h^{-2},0.6914)$ with $h= 0.6777$~\cite{Planck} are the density parameters of DM, total matter, and cosmological constant, respectively. The lifetimes $\tau_{H,L}$, masses $M_{H, L}$ and flavor weights $\epsilon^{e,\mu,\tau}_{H,L}$ for the $\gamma$-ray injection spectra of the two DMs are obtained by the $\chi^2$-fits in the previous section, which are listed in Table~\ref{tab_2DM}. As for the computation of the extragalactic IC scattering contribution, we follow the treatment in Ref.~\cite{Gamma_extraDMIC}.
The final results for the various galactic and extragalactic contributions, as well as the total $\gamma$-ray spectrum, are presented in Fig.~\ref{Fig_GammaRay}.

In Fig.~\ref{Fig_GammaRay}, different $\gamma$-ray components with different origins are evident. In the energy range $E_\gamma\lesssim  0.1~{\rm GeV}$, only the isotropic extragalactic source in Eq.~(\ref{Eq_ExtraSource}) yields the contributions. Beyond that range, the ordinary galactic background (pion decay + inverse Compton scattering + bremsstrahlung) begins to dominate the total $\gamma$-ray spectrum. The DM contributions become prominent when $E_\gamma \gtrsim 1$ GeV, extending to as high as 1500 GeV with the sharp cutoff. On top of that, the drop-out of the lighter DM component is also obvious as the fine structure around 100 GeV. Remarkably, the measured inclusive continuum $\gamma$-ray spectrum by Fermi-LAT given in Ref.~\cite{FermiLAT_Gamma} also shows this fine structure in the expected energy range. If this fine structure survives and becomes clearer with more data accumulated in the near future, it would be a strong support of the existence of the 100 GeV DM component. Moreover, our model predicts the sharp falloff above 1000 GeV, which is a clean evidence for the second DM component and could be observed by Fermi-LAT as well as other future experiments such as Cherenkov Telescope Array~\cite{CTA, CTAweb}.

Finally, it is observed from Fig.~\ref{Fig_GammaRay} that the predicted $\gamma$-ray spectrum in our model is consistent with the Fermi-LAT measurement in all the measured energy range. 
%{\color{red} 
In order to make this observation more precise, we calculate the usual $\chi^2$ with the 82 Fermi-LAT data points
and find that  $\chi^2=79.1$ with $\chi^2/d.o.f=0.965$, which is sufficient to illustrate that the  predictions in our two-component DM model 
agree with the actual measurement well.
%} 
This conclusion seems to contradict the stringent DM lifetime bound obtained 
in Ref.~\cite{Cirelli:2012ut, Papucci:2009gd}.
Here, we want to give some remarks on the possible reasons for these differences of our present result from others. The main difference of ours from Ref.~\cite{Cirelli:2012ut} lies in the interpretation of the composition of the Fermi-LAT data: Ref.~\cite{Cirelli:2012ut} assumes that the spectrum can be obtained with the conventional astrophysical sources, and can be fitted with the simple power law function. The possible contribution from DM can only be compared with the residue after the subtraction of the data points to this background function, resulting in a very stringent DM lifetime bound. In our treatment here, however, we try to calculate every $\gamma$-ray contribution precisely. Except for the EGB part from the analysis of the first-year Fermi-LAT data, the other contributions are actually already determined after we specify our CR diffusion-reacceleration parameters listed in Table~\ref{parameters} and fix the model parameters in Table~\ref{tab_2DM} by fitting the AMS-02 and Fermi-LAT data. { As for the constraint from Ref.~\cite{Papucci:2009gd}, we need to be more precise since the authors, Papucci and Strumia (PS),  
in Ref.~\cite{Papucci:2009gd} did not assume any astrophysical background at all in their derivation. 
 From Fig.~8 in Ref.~\cite{Papucci:2009gd}, we can read off the lower DM lifetime bound $\tau^{PS}_\mu = 5\times 10^{25}$s for $\mbox{DM} \to \mu^+ \mu^-$ with $m_{\rm DM}= 3000\,$GeV and $\tau_{\tau}^{PS}=1.5\times 10^{26}\,$s for $\mbox{DM}\to \tau^+\tau^-$ with $m_{\rm DM}=200\,$GeV, corresponding to the heavy and light DM dominant decaying channels with the energy cutoffs $E_{cH}=1500\,$GeV and $E_{cL}=100\,$GeV, respectively. Note that each lepton in the lepton pair decay channels carries one half of the DM energy. 
 Nevertheless, there should be a factor $1/4 = 1/2\times 1/2$ suppression in our two-component DM case, in which one 1/2 accounts for the half density of each DM component and the other 1/2 for the single $e^+$ or $e^-$ generated in one DM decay process. Also, an extra suppression from the DM mass requires to be considered. By taking all of these suppressions  into account, 
 we can transform the DM lifetime bounds shown in Ref.\cite{Papucci:2009gd} into that in our case through the following formula,
\begin{equation}
\tau_l = \frac{M^{PS}_{\rm DM} \tau^{PS}_l}{4 M_i}\,.
\end{equation}
For example, for the light DM case, the corresponding lifetime bound for the $\tau-$channel in our case is only $\tau_{\tau}= 2\times 10^{25}\,$s with the light DM mass $M_L=416\,$GeV. The same argument can be also applied to the heavy DM with the lifetime bound $\tau_\mu = 1.24\times 10^{25}\,$s. Obviously, these two bounds are much lower than the two best-fitting   DM lifetimes of $\tau_L = 8.2\times 10^{25}\,$s and $\tau_H=7.6\times 10^{25}\,$s listed in Table~\ref{tab_2DM}.}

In sum, our calculation is completely consistent with the fit of the AMS-02 positron fraction and the Fermi-LAT $e^\pm$ flux, thus representing the generic prediction of the diffuse $\gamma$-ray emission for the present multi-component decaying DM model.

\section{Microscopic Model Realization of Multi-Component Dark Matter Scenario}\label{model}
\begin{figure}
\centering
\includegraphics[width=8cm]{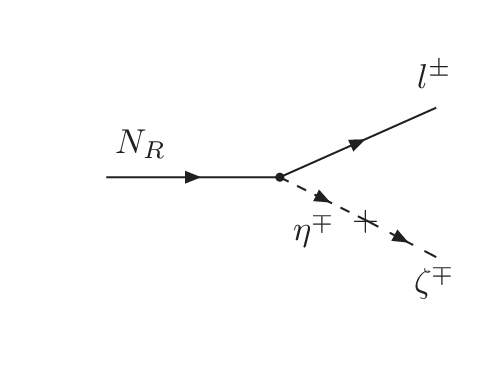}
\caption{Illustration for the process of the DM particle $N_R$ decaying into charged leptons $\l^\pm$ and heavy charged particles $\zeta^{\mp}$ through the mixings with $\eta^{\mp}$.}\label{Fig_DMDecay1}
\end{figure}

The previous phenomenological analysis has already shown that the two-component DM scenario is promising to solve the $e^+/e^-$ anomalies of the AMS-02 and Fermi-LAT data at the same time, while satisfying the diffuse $\gamma$-ray constraint from Fermi-LAT. On the other hand, a microscopic points of view would help us understand the underlying dynamics deeper. In this section, we would like to construct a particle physics model to realize this two-component DM scenario, which is a simple two-component DM extension of the one in Ref.~\cite{Chen:2009gd}. The starting point is to introduce two $SU(2)_L$ singlet fermions $N_{R\,1,2}$ with hypercharge $Y=0$ and two $SU(2)_L$ doublet scalars $\eta$ and $\zeta$ with the same hypercharge $Y=-1$. Two $Z_2$ symmetries are imposed on these newly introduced particles with the corresponding charges presented in Table~\ref{tab_z2list}.
\begin{table}
\caption{Quantum numbers of the discrete symmetries for new particles.}
\begin{tabular}[t]{rccccc}
\hline
& $N_{R\, 1}$ & $N_{R\, 2}$ & $\eta$ &$\zeta$  \\
\hline
$Z_{2}$&$-$ &$-$ & $-$  & $+$    \\
$Z'_{2}$&$+$ &$+$ & $+$  & $-$    \\
\hline
\end{tabular}
\label{tab_z2list}
\label{tab_decaymode}
\end{table}
Note that $N_{R\, 1,2}$ are the two DM candidates, achieved by requiring that the tree level mass of $\eta$ should be larger than those of $N_{R\,1,2}$ such that the decays of $N_{R\,1,2}$ through the $Z_2\times Z_2^\prime$-allowed Yukawa interactions $\bar L_{L\,i} N_{R\, 1,2} \eta$ are kinematically forbidden, where the subscripts $i=1,2,3$ stand for three generations. However, in order for the two DM components to decay, we also need to further explicitly break $Z_2\times Z_2^\prime$ by adding the soft-breaking term $\mu^2 \zeta^\dagger \eta$ with the characterizing energy scale $\mu$, as well as demanding the mass of the doublet $\zeta$ is smaller than those of two DMs $M_{1,2}$. Thus, the relevant Lagrangian to the two decaying DM components $N_{R\, 1,2}$ is given as follows:
\begin{eqnarray}
L=-\bar L_{Li} (Y_{1i}N_{R1}+ Y_{2i}N_{R2}) \eta-{M_1\over2}\overline{ (N_{R1})^c}N_{R1}
-{M_2\over2}\overline{ (N_{R2})^c}N_{R2}-\mu^2 \zeta^\dagger \eta- V\;,
\end{eqnarray}
where the scalar potential $V$ includes all other possible interactions involving $\eta$ and $\zeta$. The main decay channels of the DMs are represented in Fig.~\ref{Fig_DMDecay1}, which can be viewed as the resolution of the blob in Fig.~\ref{Fig_DMDecay}.
With this setup, the lifetimes of the two DMs of ${\cal O}(10^{26}{\rm s})$ can be naturally obtained if 
{$Y_{1(2)i}\sim{\cal O}(10^{-6})$, $\mu\sim{\cal O}(1 ~{\rm GeV})$ and $M_\eta\sim {\cal O}(10^{10} ~{\rm GeV})$.}

An interesting prediction of this model is that there is the same amount of neutrino fluxes from the DM decay as the leptons. This can be easily seen from Fig.~\ref{Fig_DMDecay1} when we make an $SU(2)_L$ rotation in terms of the decay products, $\zeta^\pm$ and $l^\mp$, to their $SU(2)_L$ neutral partners, $\zeta^0$ and $\nu_{e,\mu,\tau}$. Currently, the search for neutrinos from annihilating and decaying DMs is performed in the South Pole by the IceCube Collaboration~\cite{ICECUBE1,ICECUBE2}. The analysis of data of neutrinos from the Galactic halo~\cite{ICECUBE1} and the Galactic center~\cite{ICECUBE2} can already give quite tight bounds on the annihilation cross sections for the annihilating DM explanation of the positron/electron excesses, but for the decaying DM scenario the bound on the DM lifetime is rather weak, only from ${\cal O}(10^{22}{\rm s})$ to ${\cal O}(10^{24}{\rm s})$ for different decay channels, particularly for typical leptonic channels. As a result, our two-component model predicts that the two lifetimes of two DMs are of few$\times{\cal O}(10^{25}{\rm s})$, which can be potentially observed by the near-future IceCube experiments.

%%%%%%%%%%%%%%%%%%%%%%%%%%%%%%%%%%%%%%%%%%
\section{Conclusions}\label{conclusion}
Both the precise measurements of the positron fraction by AMS-02 and the total $e^+ + e^-$ flux by Fermi-LAT evidently show the uprise above 10 GeV, which cannot be explained by the traditional astrophysical sources. Decaying DMs could be one appealing possible origin for these extra $e^\pm$. However, for the simplest scenario with only one DM component decaying mainly through two-body leptonic channels, it is not easy to accommodate both experiments. In the present work, we have investigated the multi-component DM scenario as one possible solution to the above problem, in which at least two DM components possess their own two-body leptonic decays. As a byproduct, the fine structure around 100 GeV observed in the data of both the AMS-02 positron fraction and the Fermi-LAT total $e^+ + e^-$ flux can have the simple explanation that the lighter DM contribution drop out there. By performing the simple $\chi^2$-fitting of the two spectra, we have found that the heavier DM component with the energy cutoff larger than 1 TeV decays dominantly through the $\mu$-channel, while the lighter one with exactly 100 GeV cutoff mainly via the $\tau$-channel with the minor contribution from the direct $e$-channel.

With the fitted parameters, we have predicted the spectrum of the diffuse $\gamma$-ray emission in our two-component DM model. By comparing the spectrum with the one measured by Fermi-LAT~\cite{FermiLAT_Gamma}, we have demonstrated that it is consistent with the Fermi-LAT data points, showing that our present two-component DM model is still allowed by the current Fermi-LAT $\gamma$-ray measurement. We note that the Fermi-LAT constraint is not so stringent as pointed in the previous study~\cite{Cirelli:2012ut}.

Finally, we have built a microscopic particle model to realize the above two-component decay DM scenario. With the appropriate choice of the particle masses, mixings and couplings, it is quite natural to obtain the lifetimes of the two DM components to be of ${\cal O}(10^{26}{\rm s})$. Our scenario also predicts the same amount of the neutrino flux signal, which is expected to be observed in the future IceCube experiments.

\begin{acknowledgments}
We are grateful to Dr.~Y.~F.~Zhou and Dr.~P.~Y.~Tseng for useful discussions. The work was supported in part by National Center for Theoretical Science, National Science
Council (NSC-101-2112-M-007-006-MY3) and National Tsing Hua
University (Grant Nos. 102N1087E1 and 102N2725E1).
\end{acknowledgments}

\end{document}